\documentclass[prl,preprint,showpacs,showkeys,preprintnumbers,amsmath,amssymb]{revtex4-1}

\usepackage{graphicx}
\usepackage{dcolumn}
\usepackage{bm}

\begin{document}

\preprint{Phys. Rev. B 90, 245412 (2014)}
\title{Quantum Sticking of Atoms on Membranes}

\author{Dennis P. Clougherty}
\email{dpc@physics.uvm.edu}

\affiliation{
Department of Physics\\
University of Vermont\\
Burlington, VT 05405-0125}

\date{\today}

\begin{abstract}
A continuum model for low-energy physisorption on a membrane under tension is proposed and studied with variational mean-field theory.  A discontinuous change in the energy-dependent sticking coefficient is predicted under certain conditions.  This singularity is a result of the bosonic orthogonality catastrophe of the vibrational states of the membrane.   The energy-dependent sticking coefficient is predicted to have exponential scaling in $1/E$ above the singularity.  The application of this model to the quantum sticking of cold hydrogen to suspended graphene is discussed.  The model predicts that a beam of atomic hydrogen can be completely reflected by suspended graphene at ultralow energies.

\end{abstract}

\pacs{68.43.Mn, 03.65.Nk, 68.49.Bc, 34.50.-s}

\maketitle
\section{Introduction}
The adsorption of atoms and molecules to a surface is of fundamental interest in science. Corrosion, heterogeneous catalysis, and epitaxy all involve the adsorption of atoms to a surface in a central way. Moreover, adsorption can transform a material, since many physical properties of solids are determined in part by their adsorbates.   

The physisorption of low-energy atoms and molecules to a surface is the focus of this study.  The impetus for considering  this regime is provided by recent experimental breakthroughs in  methods to produce and manipulate ultracold atoms.  Two distinct quantum  effects  shape physisorption in this regime.  The first effect, called quantum reflection (QR), is a wave phenomenon that reduces \cite{dpc92} the likelihood of a low-energy quantum particle of reaching the surface, in comparison to a classical particle.  This reduction due to QR in the particle's probability density at low energies near the surface leads to a reduction in the transition probability of the particle to a state bound to the surface.  QR consequently leaves its imprint on the behavior of low-energy physisorption.

The second effect, the orthogonality catastrophe (OC), is a many-body effect that reduces the overlap between the state of the surface in the absence of the adsorbate and its state in the presence of the adsorbate.  The OC can cause a sharp transition in the physisorption rate  at  low energies such that one-phonon physisorption (quantum sticking) to a surface state is completely suppressed \cite{dpc11}.  While these effects have been previously considered in the context of  low-energy atomic physisorption on a semi-infinite solid, adsorption to a 2D solid such as graphene has some interesting differences from the 3D case.  

There have been a variety of experimental measurements of the sticking of cold atoms to surfaces in recent years, including, for example, systems such as: thermal neon atoms to Ru(001) surfaces \cite{neon}; spin-polarized hydrogen atoms to liquid helium surfaces \cite{doyle91,yu93}; and sodium atoms in a dilute Bose-Einstein condensate to silicon surfaces \cite{ketterle04}.  We now consider the case of the physisorption of slow atoms to membrane-like materials such as graphene.

Graphene can be suspended across $\mu$m-sized holes in a substrate, creating a high-Q nanomechanical resonator \cite{mceuen} at low temperatures.   Slow-moving adatoms would primarily exchange energy through graphene's ZA flexural \cite{graphene} modes.  These vibrations are very similar in character to those of a clamped elastic membrane under tension.  ZA flexural modes are polarized normal to the membrane in equilibrium.  In contrast, the surface vibrations  of a 3D elastic solid have a different character with only partial polarization normal to the surface.  Thus the inelastic atom-membrane coupling might be enhanced relative to a 3D solid.  Furthermore, the vibrational spectrum of suspended graphene depends on the membrane tension and is independent of elastic constants.  Thus, the inelastic atom-membrane coupling is in principle a parameter that might be tuned experimentally.

A recent numerical study \cite{graphene} of the atomic sticking to graphene concluded that at low incident energies the sticking probability of atomic hydrogen to suspended graphene is enhanced relative to a graphite surface.  The model used in this numerical study however suffers within perturbation theory from a divergent self-energy as a result of the low-frequency behavior of the atom-phonon  \cite{dpc12} interaction.  A low-frequency cutoff was used to make the numerical calculation tractable \cite{LJcutoff}, and the effects of the infrared divergence were not contained in their results.  A  continuum version of their model with the same low-frequency behavior in the atom-phonon interaction is considered here in an effort to explore the consequences of this infrared divergence on physisorption.  Such a model should also apply to the growing number of membrane-like 2D materials  (boron nitride, silicon nitride \cite{sin} and the monolayer transition metal dichalcogenides serve as examples of these ``quantum drums'' \cite{drum}) available to experiment.

\section{Atom-Membrane Interaction}

One might  start by treating the atom-membrane interaction to be a sum of two-body interactions over the surface of the membrane.  
For an atom located a distance $Z$ above the membrane center, the atom-membrane interaction is of the form
\begin{equation}
V(Z)=\int d^2 r\ v(\sqrt{r^2+(Z-u)^2})
\end{equation}
where $v(\sqrt{r^2+(Z-u)^2})$ is the effective interaction between a differential patch of the membrane located at $({\bf r}, u)$ and the impinging atom at $Z$.  (Here, $u$ is the height of the patch above the $xy$-plane and ${\bf r}$ sweeps over the membrane in its  equilibrium in isolation.)  Expanding the interaction to linear order in $u$, one obtains
\begin{equation}
V(Z)=\int d^2 r v(\sqrt{r^2+Z^2})-{\partial\over\partial Z}\int d^2 r  u({\bf r}) v(\sqrt{r^2+Z^2})
\label{pot}
\end{equation}
The interest here is in low energies and low temperatures where the validity of neglecting higher order terms in the displacement to describe atom-surface collisions has been previously discussed \cite{gumhalter03}.

The first term in Eq.~\ref{pot} is the interaction with a static, flat membrane.  For the van der Waals case, the long-range attractive interaction between two neutral, polarizable particles separated by $s$ behaves as $v(s)=-C_6 s^{-6}$ for large $s$ (neglecting retardation effects).  Thus, the static term behaves at large distances as
\begin{equation}
V_0(Z)=-{\pi C_6\over 2}\bigg({1\over Z^4}-{1\over (Z^2+a^2)^2}\bigg)
\label{static}
\end{equation}
for a circular membrane with radius $a$.  This is asymptotically equal to the Casimir-Polder potential between a neutral atom and a 2D insulating solid \cite{dobson}.  The short-range repulsive contribution to the atom-membrane interaction can be similarly obtained. 

For incident atoms approaching near the membrane's center, the static interaction can be expanded efficiently in cylindrical multipole moments.  Thus for off-axis collisions the  interaction of Eq.~\ref{static} is the lowest-order term in an expansion of $R/a$ where $R$ is the distance of the atom from the membrane's symmetry axis.

\begin{figure}
\includegraphics[width=15cm]{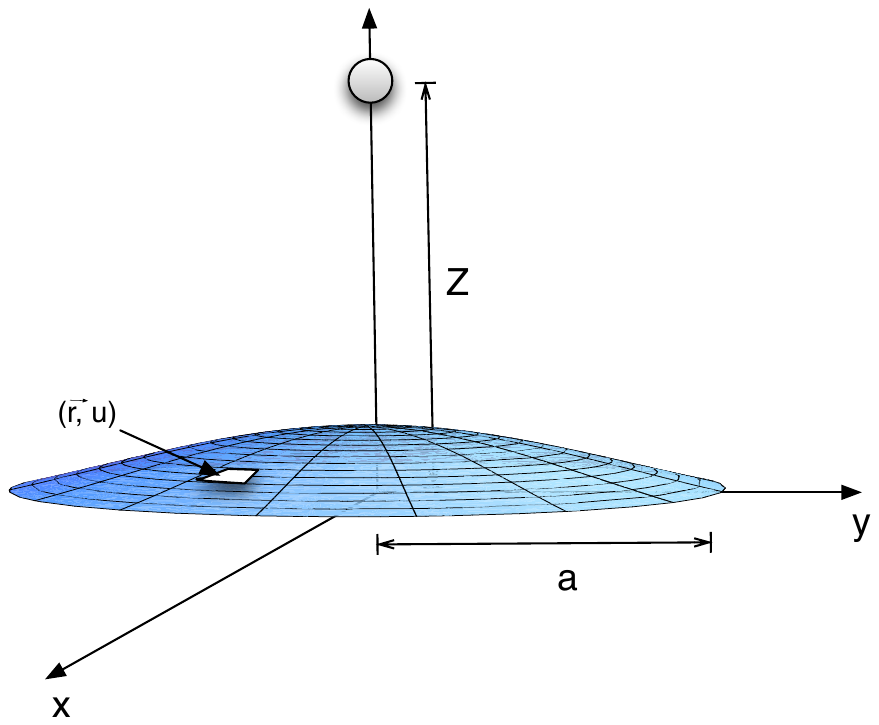}
\caption{\label{fig:membrane}  Sketch of the circular membrane radius $a$ with an impinging atom mass $M$. The membrane distorts out of the xy-plane in the presence of the adatom. Each differential patch of the membrane, located at $(r, \theta, u)$ (polar coordinates), contributes to the atom-membrane interaction.}
\end{figure}

The second term in Eq.~\ref{pot} is the inelastic atom-membrane interaction, coupling the atom to excitations of the membrane.  Classically, the sound that a drum makes depends on the location of the strike on the drumhead.  For an impulse directed at the drum's center, only circularly symmetric modes are excited.  Similarly, it will be shown for an atomic beam focussed on the center of suspended graphene, only the circularly symmetric modes participate in the inelastic scattering.  It will be apparent that only this portion of graphene's vibrational density of states is relevant to the sticking process under the assumed conditions.

One can expand the displacement $u$ in normal modes of the clamped membrane \cite{mf} $\rho_{mn}({\bf r})$
\begin{equation}
u({\bf r})=\sum_{m=-\infty}^\infty \sum_{n=1}^\infty Q_{mn} \rho_{mn}({\bf r})
\end{equation}
where
\begin{equation}
\rho_{mn}({\bf r})={1\over\sqrt{\pi a^2}}{J_m({\alpha_{mn} r/ a})\over |J_{m+1}(\alpha_{mn})|}e^{im\theta}
\label{modes}
\end{equation}
and $\alpha_{mn}$ is the nth root of $J_m$.  
Hence,  the inelastic interaction at large distances becomes
\begin{eqnarray}
V_1&=&-\sum_{m=-\infty}^\infty \sum_{n=1}^\infty Q_{mn} {\partial\over\partial Z}\int d^2 r  \rho_{mn}({\bf r}) v(\sqrt{r^2+Z^2})\cr
&=&{C_6\over\sqrt{\pi a^2}}\sum_{m, n} {Q_{mn}\over |J_{m+1}(\alpha_{mn})|} {\partial\over\partial Z}\int d\theta dr r  {J_m({\alpha_{mn} r/ a})\over(r^2+Z^2)^3} e^{im\theta}\cr
&=&{2\pi C_6\over\sqrt{\pi a^2}}\sum_{n} {Q_{0n}\over |J_{1}(\alpha_{0 n})|} {\partial\over\partial Z}\int  dr r  {J_0({\alpha_{0 n} r/ a})\over(r^2+Z^2)^3} 
\end{eqnarray}
Thus, only the $m=0$ modes participate in the scattering.
For $a>Z$, $V_1$ becomes
\begin{equation}
V_1\approx {2\pi C_6\over\sqrt{\pi a^2}}\sum_{n} {Q_{0n}\over |J_{1}(\alpha_{0 n})|}  {\partial\over\partial Z} \bigg({k_{0n}^2\over 8 Z^2} K_2(k_{0 n}Z)\bigg)
\end{equation}
where $k_{m n}=\alpha_{m n}/a$ and $K_2$ is a modified Bessel function of the second kind.

By quantizing the vibrations of the membrane (see Appendix), one obtains 
\begin{equation}
Q_{mn}=\sqrt{\hbar\over 2\sigma\omega_{mn}} (b_{mn}+b^\dagger_{\bar m n})
\end{equation}
where $\omega_{mn}= v_s k_{m n}$ and $b^\dagger_{mn}$ ($b_{mn}$) is a creation (annihilation) operator for quanta in the mode labeled by $(m, n)$.  The speed of sound is determined by  the membrane tension $\gamma$,  and the membrane mass density $\sigma$, viz. $v_s=\sqrt{\gamma\over\sigma}$.   Thus,
\begin{equation}
V_1\approx {2\pi C_6\over\sqrt{\pi a^2}}\sum_{n}\sqrt{\hbar\over 2\sigma\omega_{0n}} {1\over |J_{1}(\alpha_{0 n})|}  {\partial\over\partial Z} \bigg({k_{0n}^2\over 8 Z^2} K_2(k_{0 n}Z)\bigg) (b_{0n}+b^\dagger_{0 n})
\end{equation}
For notational simplicity, the $m=0$ subscript is dropped in what follows.

\section{Hamiltonian}

One can truncate the atom state space to the continuum state $|k\rangle$ initially occupied  and the bound state $|b\rangle$ in the static potential $V_0$.  The Hamiltonian is then of the form \cite{dpc11}
\begin{equation}
H=H_p+H_b+H_c
\label{ham}
\end{equation}
where
\begin{eqnarray}
H_p&=&E c_k^\dagger c_k -E_b c_b^\dagger c_b,\\
H_b&=&\sum_n{\hbar\omega_n {b_n^\dagger} b_n},\\
H_c&=&-(c_k^\dagger c_b+c_b^\dagger c_k)g_{kb}\sum_n \xi\ ({b_n+b_n^\dagger}) 
-c_k^\dagger c_k g_{kk}\sum_n \xi\ ({b_n+b_n^\dagger}) \nonumber\\
&&- c_b^\dagger c_b g_{bb}\sum_n \xi \ ({b_n+b_n^\dagger}) 
\label{gbb}
\end{eqnarray}
and $\xi=\sqrt{\hbar \over 4a\sigma v_s}$ with $g_{\alpha\beta}=\langle \alpha|V_0'(Z)|\beta\rangle$.

The frequency independence of $\xi$, the coupling of the atom to low frequency phonons,  in this model leads to problems with a straightforward perturbative expansion: the atom-phonon coupling will shift the binding energy of the atom to the membrane, and a calculation of the atom self-energy to second-order in $g_{bb}$  logarithmically diverges with increasing membrane size \cite{dpc12}. Thus, results based on finite-order perturbation theory are unreliable for this model.  

\section{Variational mean-field theory}

In previous work \cite{dpc10, dpc11} a variational mean-field method has been used to obtain the sticking probability at low energies.  A generalized unitary transformation $U=\exp(c_b^\dagger c_b x)$, $x\equiv \sum_n {{f_n} (b_n-b_n^\dagger)/\hbar\omega_n}$ that displaces the membrane when in the presence of the adatom is applied to the Hamiltonian $H$, resulting in the  transformed Hamiltonian $\tilde H=H_0+H_1$, 
\begin{equation}
H_0=E c_k^\dagger c_k -\epsilon c_b^\dagger c_b-\Delta c_b^\dagger c_k-\Delta^* c_k^\dagger c_b+
\sum_n{\hbar\omega_n {b_n^\dagger} b_n}
\end{equation}
\begin{eqnarray}
H_1&=& -c^{\dagger}_{k}c_b\left(\sum_{n}g_{kb}\xi(b_{n}+b_{n}^{\dagger})e^{-x}-\Delta^{*}\right)-c_b^{\dagger}c_{k}\left(e^{x}\sum_{n}g_{kb}\xi(b_{n}+b_{n}^{\dagger})-\Delta\right)\nonumber\\
&&-c_{k}^{\dagger}c_{k}\sum_{n}g_{kk}\xi(b_{n}+b_{n}^{\dagger})-c_b^{\dagger}c_b\sum_{n}(g_{bb}\xi-f_{n})(b_{n}+b_{n}^{\dagger})
\label{interaction}
\end{eqnarray}
where
\begin{eqnarray}
\epsilon&\equiv& E_b+\Delta E_b\\
\Delta E_b&=& \sum_{n}\frac{2f_{n}g_{bb}\xi-f_{n}^{2}}{\hbar\omega_{n}}
\end{eqnarray}
and 
\begin{equation}
\Delta\equiv \left\langle e^{x}\sum_{n}g_{kb}\xi\ (b_{n}+b_{n}^{\dagger})\right\rangle
\label{delta}
\end{equation}
$ \langle\cdots\rangle$ denotes the thermal average over the phonon states.  $\Delta E_b$ is a shift in the bound state energy of the atom that results from the atom-phonon interaction.

The optimum values of the parameters of the transformation $\{f_n\}$ are determined by minimizing the Bogoliubov-Peierls upper bound to the free energy of the system  \cite{sb, dpc11}.  One thus obtains the following expression for the variational parameters, valid for $E_b \gg \Delta$
\begin{equation}
f_{n}=\frac{g_{bb}\ \xi}{1+\frac{\bar\omega}{\omega_{n}}\coth\frac{\beta\hbar\omega_{n}}{2}}
\label{fns}
\end{equation}
where $\hbar\bar\omega\equiv\Delta^2/(E+\epsilon)$ and $\beta^{-1}=k_B T$.

From Eq.~\ref{fns} one concludes that if the frequency of the nth mode $\omega_n$ is large with respect to  $\bar\omega$, then $f_n\approx g_{bb}\ \xi$.  Thus, in this case it is apparent from Eq.~\ref{interaction} that $U$ fully eliminates the interaction with the nth mode when the atom is bound to the membrane.  For low frequency modes where $\omega_n\ll\bar\omega$, Eq.~\ref{fns} gives $f_n$ vanishing as $\omega_n^2$.  Thus the fast, high-frequency modes yield a new equilibrium  for the membrane in the presence of the impinging atom that is distorted relative to that of the membrane in isolation.  This is reminiscent of Leggett's adiabatic renormalization \cite{leggett} in the spin-boson model.

\section{Self-consistent solution}

A self-consistent equation for the mean-field transitional amplitude $\Delta$ may be obtained from Eqs.~\ref{delta} and \ref{fns} within the continuum approximation:
\begin{equation}
\Delta\approx \delta {g_{kb}\over g_{bb}} \ln{ \epsilon_D \epsilon\over\Delta^2} \exp(-{\epsilon \delta \over \Delta^2})
\label{sce}
\end{equation}
where $\rho= a/\pi \hbar v_s$ is the density of vibrational modes (independent of frequency in this model), $\epsilon_D\equiv\hbar\omega_D$ is the energy of the highest frequency phonon supported by the membrane, and $\delta\equiv\rho g_{bb}^2\xi^2$.  Eq.~\ref{sce} applies to the low temperature case where the membrane temperature satisfies $k_B T\ll \delta$.

\begin{figure}
\includegraphics[width=12cm]{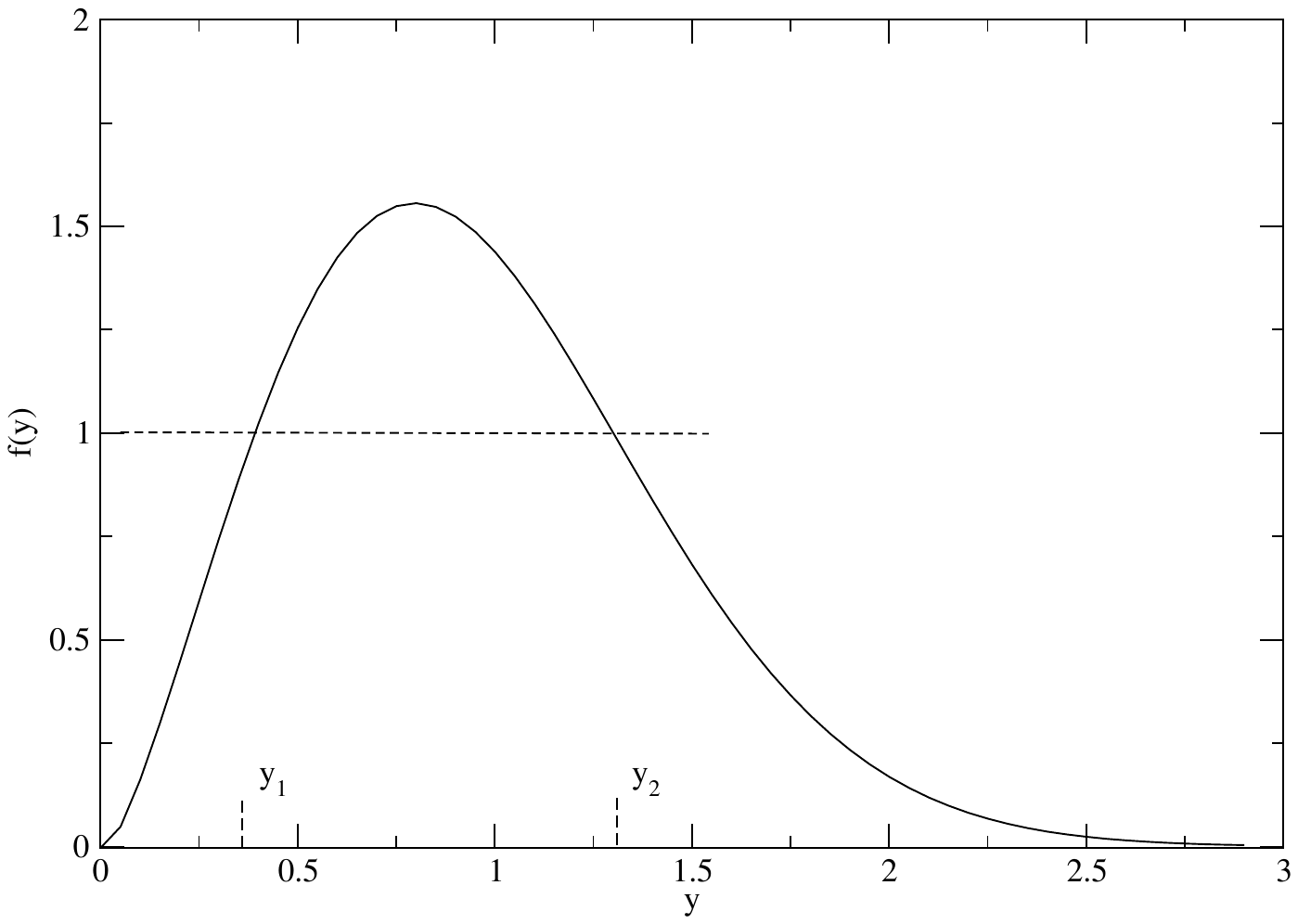}
\caption{\label{fig:sce}  Graphical solution of the self-consistent equation for $\Delta$ for the case of $C=1$ and $A=50$.  The solution $y=y_1$ corresponds to larger value of $\Delta$ and the minimum free energy. }
\end{figure}

Eq.~\ref{sce} can be solved graphically (Fig.~\ref{fig:sce}). By setting $y=\sqrt{{\epsilon \delta /\Delta^2}}$, the self-consistent condition reduces to $f(y)=1/C$ where $f(y)\equiv y\ln A y \exp(-y^2)$ and 
the dimensionless parameter $C$ is given by
$C\equiv{2 g_{kb}\over g_{bb}} \sqrt{\delta\over \epsilon} $
and $A\equiv\sqrt{\epsilon_D/\delta}$.
The value of $C$ depends on the incident energy $E$ through $g_{kb}$.

From Eq.~\ref{sce}, it is apparent that $\Delta=0$ is always a solution.    In addition, a non-vanishing positive, real solution for $\Delta$ exists when $C=C_*(\approx \sqrt{2 e}/\ln (A/\sqrt{2}))$ for $A\gg 1$.   (Here, $e$  is Napier's constant, $e=2.71828\dots$.)  For $C>C_*$, two positive, real solutions for $\Delta$ exist; for $C< C_*$, only the trivial solution $\Delta=0$ exists.  

It is straightforward to show that the largest value of $\Delta$ possible yields the minimum free energy.  Hence, $\Delta$ changes discontinuously at criticality, dropping from $\Delta=\Delta_c(\approx \sqrt{2 \epsilon\delta})$ for $C=C_*$ to $\Delta=0$ for $C< C_*$.  

The form of the self-consistent equation in Eq.~\ref{sce} is substantially different from the non-perturbative model previously considered for a 3D target \cite{dpc11} where $\Delta$ was found to smoothly vanish at a critical incident energy.  The difference is due to the low-frequency enhancement of the atom-phonon coupling in the case of the membrane.

\section{Quantum sticking rate}

The rate of transitions from the continuum state with incident energy $E$ to the lowest energy bound state can be calculated using Fermi's golden rule 
\begin{equation}
\Gamma={2\pi\over\hbar}\sum_{i f} p_i \left|\left\langle b;f\left|\tilde{H}_{int}\right|k;i\right\rangle\right|^{2}\delta\left(E+\epsilon-\Delta\epsilon_{fi}\right)
\label{rate}
\end{equation}
with 
\begin{eqnarray}
\tilde H_{int}&=&-c_k^\dagger c_b g_{kb}\sum_n \xi\ ({b_n+b_n^\dagger}) e^{-x}
-c_b^\dagger c_k e^x g_{kb}\sum_n \xi\ ({b_n+b_n^\dagger})
\end{eqnarray}
and $p_i\propto\exp(-\beta\epsilon_i)$, the probability of  the membrane initially having vibrational energy $\epsilon_i$ and $\Delta\epsilon_{fi}=\epsilon_f-\epsilon_i$.  
An incident beam with uniform circular cross section (radius $R_0\ll a$) is the assumed initial state $|k\rangle$.

The sticking rate is proportional to the square of the Franck-Condon factor $\Gamma\propto\exp(-2 F)$ where
\begin{equation}
2 F=\sum_n {f_n^2\over(\hbar\omega_n)^2} \coth {\beta\hbar\omega_n\over 2}
\label{fc}
\end{equation}
For the case of low incident energy $E<E_c$, $g_{kb}$ becomes too small to support a non-vanishing solution to Eq.~\ref{sce} ($C< C_*$).  With $\Delta=0$, the Franck-Condon factor tends to vanish for large membranes, with
\begin{equation}
e^{-2 F}\sim \exp\bigg(-{ k_B T  \delta \over \epsilon_0^2}\bigg),\ \ \epsilon_0\to 0
\end{equation}
where $\epsilon_0$ is the energy of the lowest frequency phonon supported by the clamped membrane.   As the radius $a$ becomes large, $\epsilon_0$ tends to zero in inverse proportion to $a$. Thus, the sticking rate of low-energy particles is exponentially suppressed for large membranes.

For incident energies approaching $E_c$ from above, Eq.~\ref{fc} yields $2F\approx 2\delta/\hbar\bar\omega$.  For low incident energies, $g_{kb}\propto\sqrt E$ and $\hbar\bar\omega\propto E$, an artifact of the energy scaling due to quantum reflection \cite{dpc10,dpc11}.  Thus,
the  Franck-Condon factor contributes $\exp(-E_c/E)$ to the sticking rate.
\begin{figure}
\includegraphics[width=14cm]{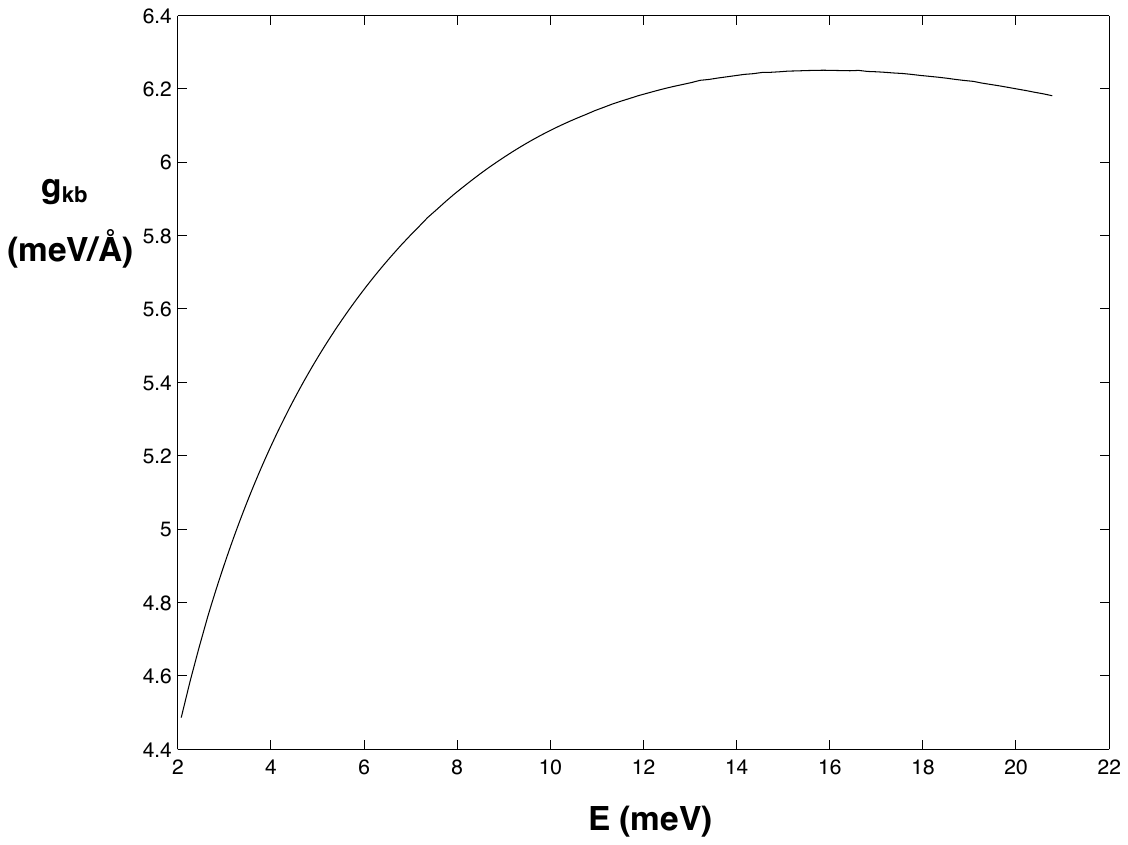}
\caption{\label{fig:gkb}  Plot of $g_{kb}$ versus incident energy $E$ for a hydrogen atom interacting with graphene.  Energy scaling is consistent with quantum reflection in the sub-meV range, with $g_{kb}\propto\sqrt{E}$.  The maximum value of $g_{kb}\approx 6.25$ meV/$\AA$, found near $E\approx 16$ meV, is roughly a factor of six below the critical value necessary to give a non-vanishing $\Delta$.}
\end{figure}

The total sticking rate is a sum over contributions from all bound states.  Thus at low temperatures such that $\epsilon_0\ll k_B T \ll \delta$,
\begin{eqnarray}
\Gamma&=&\sum_{n=1}^{n_b} \Gamma_n\cr
&\approx&{2\pi\delta\over\hbar} \bigg({g_{kb}\over g_{bb}}\bigg)^2 e^{-2 F} \sum_{n=1}^{n_b} {2\pi\over\alpha_{0n}} J^2_1({\alpha_{0n} R_0\over a})
\label{gamma}
\end{eqnarray}
where $n_b$ is the radial quantum number of the highest bound state and $R_0$ is the radius of the incident atomic beam.  The sum in Eq.~\ref{gamma} approaches a number close to unity as $n_b$ becomes large compared to $a/R_0$.

The probability of sticking is the rate of sticking per incoming flux of atoms.  It is straightforward to show from Eq.~\ref{gamma} that for energies above $E_c$, the sticking probability behaves as
\begin{equation}
s\propto\sqrt{E} e^{-E_c/E}
\end{equation}
The exponential scaling in $1/E$ is different from the case of sticking to an elastic 3D solid \cite{dpc11} where the sticking above the superreflective transition behaves as $s\propto\sqrt{E} e^{-\sqrt{E_c/E}}$.  The difference is a direct result of the difference in the low-frequency behavior of the inelastic atom-surface interaction; the inelastic atom-membrane interaction is enhanced at low frequencies in comparison to the interaction for a 3D solid.

\section{Results for graphene}
For the case of suspended graphene, potential parameters are obtained from a comparison of the asymptotic behavior of Eq.~\ref{pot} to a previous model \cite{graphene, jackson}. Numerical calculations give a binding energy of $\epsilon\approx 25$ meV and $\delta\approx 60\ \mu$eV.
The ratio of the  binding energy $\epsilon$  to $\delta$ is sufficiently large that $C$ remains much smaller than $C_*$ over  incident energies less than 15 meV for atomic hydrogen.  Thus, $E_c$ cannot be less than 15 meV for this system.  In this case, there is only the self-consistent solution $\Delta=0$ for $E\lesssim 15$ meV.  One concludes that within this model, quantum sticking is forbidden for ultracold atomic hydrogen impinging on graphene under the conditions considered.  Graphene might then serve as a perfect atomic mirror in this regime, completely reflecting incident matter waves.  The model suggests that other 2D solids with  large values of $\epsilon/\delta$  are  also potential candidates for low-loss atomic mirrors.

\begin{figure}
\includegraphics[width=14cm]{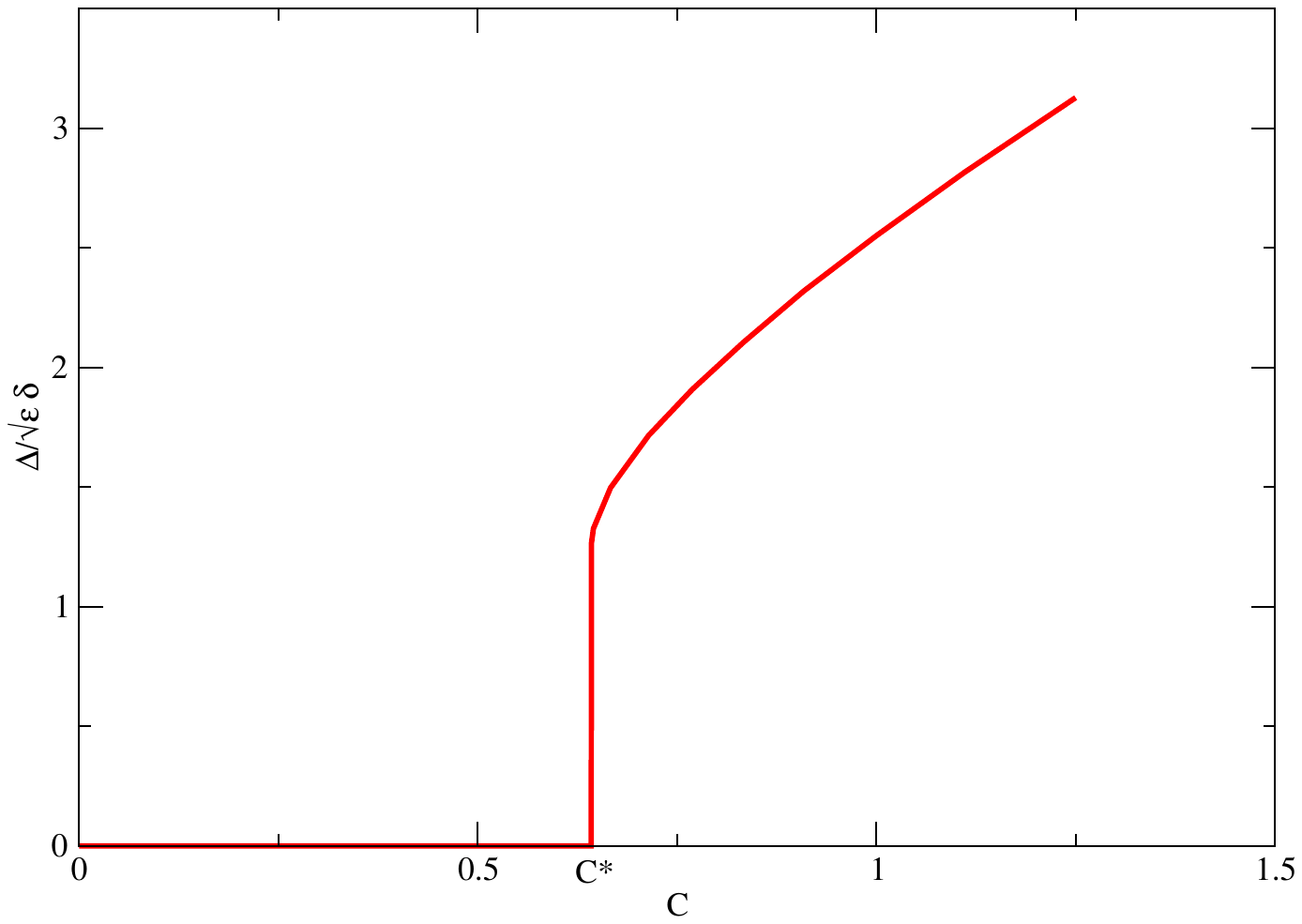}
\caption{\label{fig:delta}  Plot of $\Delta/\sqrt{\epsilon\delta}$ versus $C$ for $A=50$. Here, $C_*\approx \sqrt{2e}/\ln (A/\sqrt{2})\approx 0.65$ and $\Delta_c/\sqrt{\epsilon\delta}\approx\sqrt{2}$.}
\end{figure}

\section{Summary}
A continuum model for low-energy   sticking of a quantum particle to a membrane was proposed and studied with a variational mean-field method.  A discontinuous change in the energy-dependent sticking coefficient is found at a critical energy $E_c$.  This discontinuity is a result of low-frequency fluctuations of the membrane that  suppress the sticking rate for particles with energies below $E_c$.  The energy-dependent sticking coefficient is predicted to have exponential scaling in $1/E$ above the discontinuity.  

This model is then applied to the case of the sticking of cold hydrogen to suspended graphene.  Because of the large binding energy of hydrogen to graphene (relative to $\delta$),  the model predicts that atomic hydrogen is completely reflected by suspended graphene in the quantum sticking regime where hydrogen energies are below 15 meV.  

These non-perturbative results are contrary to a recent numerical calculation \cite{graphene} over the energy range of 1-20 meV where the phonon Fock space is restricted to zero and one-phonon states and  a low-frequency cutoff is used to make the numerics tractable.  In this restricted subspace,  one would not expect that the effects of the infrared divergence should appear, as the use of a low-frequency cutoff prevents the divergence of $F$ in Eq.~\ref{fc}.

Support by the National Science Foundation (Grants No. DMR-0814377 and No. DMR-1062966) is gratefully acknowledged. 
\vfil
\eject

\bibliography{qs}

\vfil
\eject

\appendix*
\section*{Quantization of the Vibrations of a Membrane}

The continuum Lagrange density for a membrane under tension is
\begin{equation}
\label{lagrangedens}
\mathcal{L}=\frac{1}{2}\sigma \dot{u}^{2}-\frac{1}{2}\gamma|\nabla u|^{2}
\end{equation}

The corresponding Hamiltonian is thus
\begin{equation}
\label{ham1}
H=\int \mathcal{H} d^2 r =\int  (\frac{1}{2\sigma} \Pi^{2}+\frac{1}{2}\gamma|\nabla u|^{2}) d^2 r
\end{equation}
where $\Pi$ is the canonical momentum density.

The normal displacement field $u$ can be expanded in normal modes  of clamped membrane ($u(a)=0$).

\begin{equation}
u({\bf r})=\sum_{m=-\infty}^\infty \sum_{n=1}^\infty Q_{mn} \rho_{mn}({\bf r})
\label{displace}
\end{equation}
where the (normalized)  modes of the membrane $\rho_{mn}({\bf r})$ are given in Eq.~\ref{modes}.

The momentum density can also be expanded in normal modes
\begin{equation}
\Pi({\bf r})=\sum_{n=1}^\infty \sum_{m=-\infty}^\infty P_{\bar m n} \rho_{mn}({\bf r})
\end{equation}

The Hamiltonian is thus
\begin{equation}
H=\sum_{n, m} (\frac{1}{2\sigma} P_{mn} P_{\bar m n}+\frac{\sigma}{2} \omega^2_{mn} Q_{mn} Q_{\bar m n})
\end{equation}

Imposing the quantization condition $[u({\bf r}), \Pi({\bf r'})]=i\hbar \delta({\bf r}-{\bf r'})$ and introducing the creation and annihilation operators in this polar basis
\begin{eqnarray}
b_{mn}&={i}\sqrt{1\over 2\sigma\hbar\omega_{mn}}P_{\bar m n}+\sqrt{\sigma\omega_{mn}\over 2\hbar}  Q_{mn}\cr
b^\dagger_{mn}&=-{i}\sqrt{1\over 2\sigma\hbar\omega_{mn}}P_{mn}+\sqrt{\sigma\omega_{mn}\over 2\hbar} Q_{\bar m n}
\label{creation}
\end{eqnarray}
yields the following Hamiltonian 
\begin{equation}
H=\sum_{n, m} \hbar\omega_{mn}(b^\dagger_{mn} b_{mn}+\frac{1}{2})
\end{equation}

The displacement $u({\bf r})$ can then be expressed in second quantized form using Eqs.~\ref{displace} and \ref{creation}
\begin{equation}
u({\bf r})=\sum_{m=-\infty}^\infty \sum_{n=1}^\infty \sqrt{\hbar\over 2\sigma \omega_{mn}} \rho_{mn}({\bf r}) (b_{mn}+b^\dagger_{\bar m n})
\end{equation}

\end{document}